\documentclass[
    ,final            
  ]
  {aipproc}

\layoutstyle{6x9}

\begin{document}

\title{Inclusive Hadron Production in p+p Collisions at STAR}

\classification{13.85.Ni, 13.87.Fh}
\keywords      {}

\author{Frank Simon (for the STAR collaboration)}{
  address={Massachusetts Institute of Technology}
}

\begin{abstract}
The STAR experiment at RHIC has measured a variety of inclusive hadron cross sections in $p+p$ collisions at $\sqrt{s}$ = 200 GeV. Measurements of the differential cross section for inclusive charged pion production at mid rapidity and for inclusive neutral pion production at forward rapidity ($3.0 < \eta < 4.2$) as well as the first preliminary result from STAR for the differential cross section for inclusive neutral pion production near mid rapidity are presented. These cross sections are compared to next-to-leading order perturbative QCD calculations and can provide constraints on the pion fragmentation functions. Good agreement between data and pQCD has been found for all three cross sections.

\end{abstract}

\maketitle

\vspace{-10mm}

\section{Introduction}

One goal of the polarized p+p program at the Relativistic Heavy Ion Collider (RHIC)
is the study of the gluon polarization $\Delta G$ in the
proton via spin asymmetry measurements in a variety of processes \cite{Bunce:2000uv}. Such studies typically involve fits to polarized data using perturbative QCD (pQCD) evaluation for the various processes. 
In order to validate the application of pQCD a good understanding of the (unpolarized) cross sections of these processes is crucial. In addition the study of inclusive cross sections of identified particles provides important information on the general applicability of pQCD calculations in the RHIC energy regime as well as on fragmentation functions.

The STAR experiment at the RHIC \cite{Ackermann:2002ad} is well suited for inclusive hadron measurements over a wide range in rapidity due to its central large acceptance time projection chamber (TPC) and electromagnetic calorimeters, a time of flight (TOF) detection system for particle identification and the forward neutral pion detectors (FPD).

\vspace{-2mm}
\section{Experiment and Data Analysis}

The forward neutral pion cross section is measured with the FPDs \cite{Adams:2003fx, Adams:2006uz}. These detectors are high resolution electromagnetic calorimeters consisting of lead glass crystals, covering the pseudorapidity range $\sim 3.0 < \vert\eta\vert< \sim 4.2$ on both sides of the interaction region. Events were triggered by a minimum energy requirement in the FPD in addition to the $p+p$ minimum bias trigger based on a coincidence of the beam beam counters (BBC). Neutral pions are reconstructed by reconstructing the invariant mass from their two decay photons. The mass peak position is used to calibrate the FPD energy scale to a precision of $\sim$1\%. The data presented here were taken in the 2002 and the 2003 RHIC $p+p$ run, with an integrated luminosity of 350 nb$^{-1}$.  The cross section is determined in three ranges with mean pseudorapidities $\left< \eta \right>$ = 3.3, 3.8 and 4.0.

Charged pions are identified using the TPC and the TOF system. The TPC tracks charged particles in the pseudorapidity range $-1 < \eta < 1$. For the 2003 RHIC run a first prototype of the TOF was installed and is used to provide pion-kaon separation below $p_t$ = 2.5 GeV \cite{Adams:2003qm}. Above that threshold, the ionization energy loss $dE/dx$ in the TPC is used to identify pions. Due to the relativistic rise of the energy loss, the pion $dE/dx$ is $\sim 15\%$ higher than that of kaons and can be used to separate the two particle species. A detailed description of the charged pion analysis is given in \cite{Adams:2006nd}.

\begin{figure}
\begin{minipage}[t]{.529\textwidth}
  \includegraphics[width=.99\textwidth]{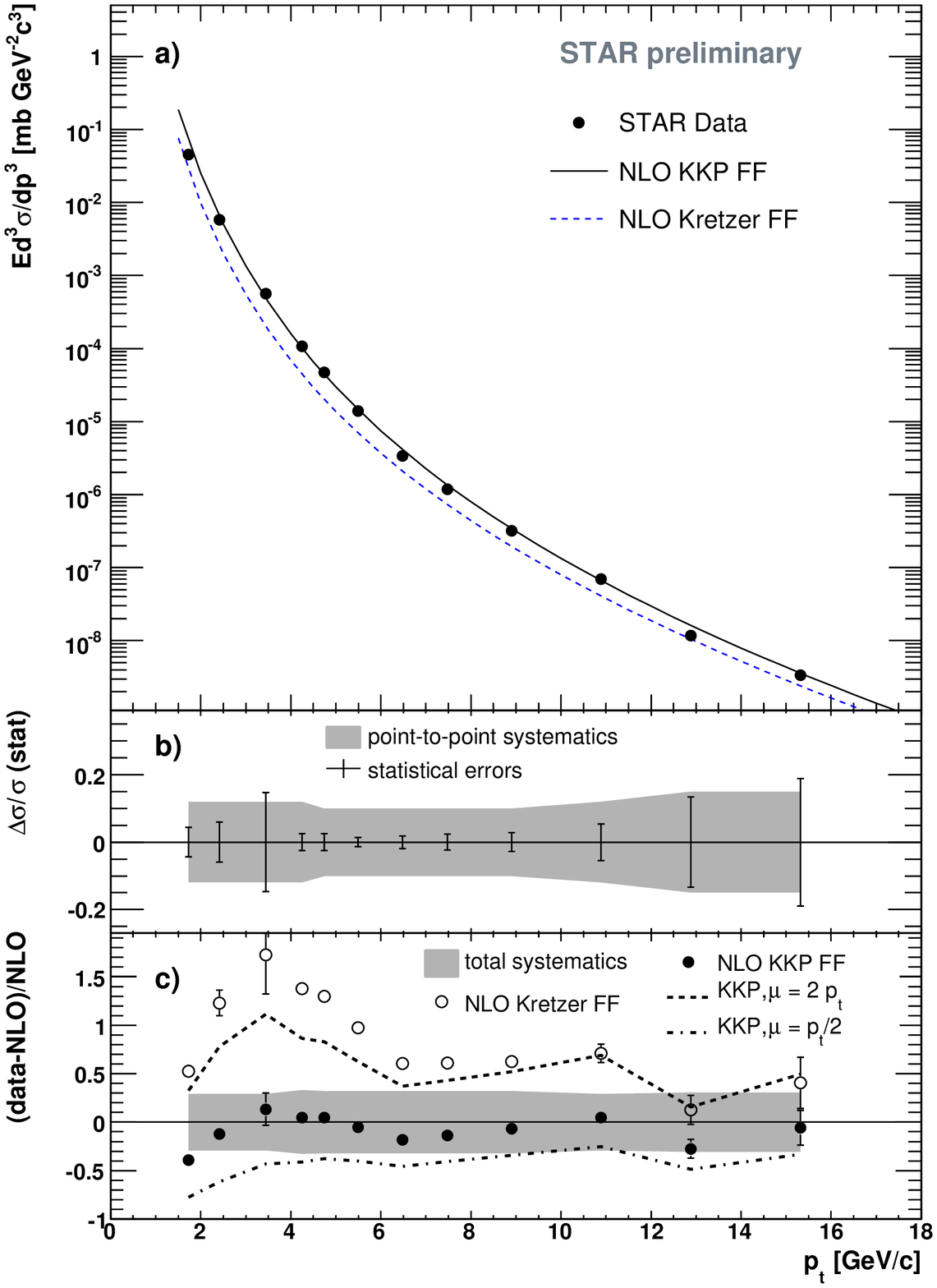}
\end{minipage}

\begin{minipage}[t]{.479\textwidth}
  \includegraphics[width=.99\textwidth]{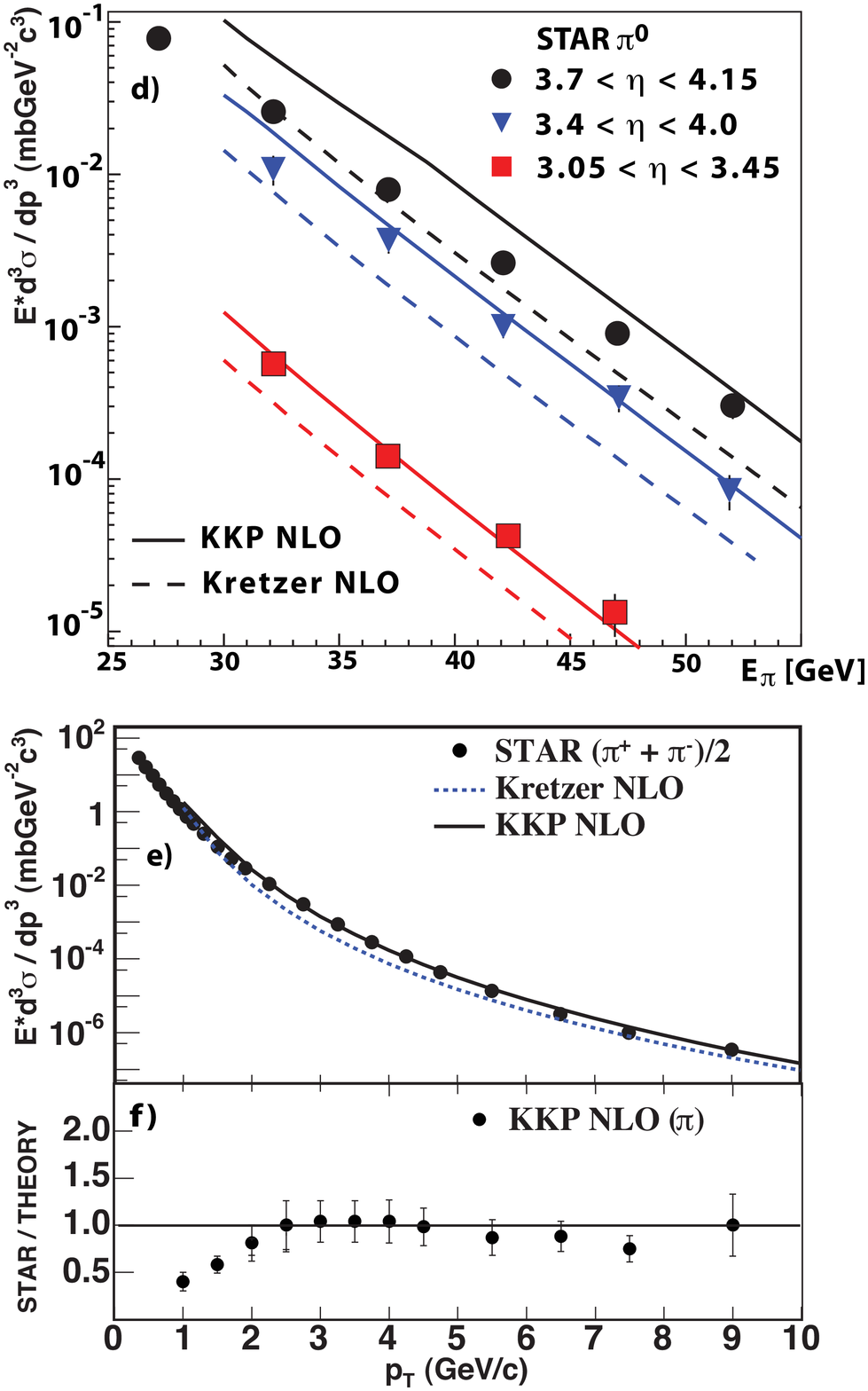}
 \end{minipage}
\caption{Inclusive hadron cross sections in p+p collisions at $\sqrt{s}$ = 200 GeV. \newline 
{\bf Left:} Mid rapidity $\pi^0$ cross section: a) compared to NLO pQCD calculations with two sets of fragmentation functions; b) relative statistical error (bars) and point-to-point systematics (band); c) relative difference to NLO pQCD calculations. For the KKP fragmentation function, the scale uncertainty is indicated by the dotted ($\mu$ = 2 p$_{\mbox{t}}$) and dashed-dotted ($\mu$ =  p$_{\mbox{t}} / 2$) lines. \newline
{\bf Right:} d) Forward rapidity $\pi^0$ cross section for three rapidity intervals \cite{Adams:2006uz}, compared to NLO pQCD calculations. e) Mid rapidity charged pion cross section \cite{Adams:2006nd}, compared to NLO pQCD with two sets of fragmentation functions. f) Ratio with theory for the KKP fragmentation function.}
\label{fig:XSect}
\end{figure}

Neutral pions near mid-rapidity are detected with the barrel electromagnetic calorimeter BEMC \cite{Beddo:2002zx}. For the 2005 run period, half of the BEMC was fully installed and commissioned, giving a coverage of $0 < \eta < 1$ for all azimuthal angles $\varphi$. The BEMC is a lead scintillator sampling calorimeter with a granularity of $0.05 \times 0.05$ in $\Delta \eta \times \Delta \varphi$. For the reconstruction of high $p_t$ neutral pions the identification of the two decay photons with a small opening angle is crucial. This is achieved using the shower maximum detector SMD, a wire proportional counter with cathode strip readout at $\sim 5\, X_0$ depth in the calorimeter modules which provides a segmentation of $0.007 \times 0.007$ in $\Delta \eta \times \Delta \varphi$. Neutral pions are identified from the $\gamma \gamma$ invariant mass spectrum. The mass and $p_t$ are determined from the opening angle between the two photons measured with the SMD and from the neutral energy deposited in the calorimeter. A charged particle veto rejects calorimeter hits that have contributions from charged particles. Dedicated online triggers, selecting events with high energy deposit in a BEMC tower in addition to the minimum bias trigger condition, were used to select events containing high $p_t$ neutral pions.  These high tower (HT) triggers used thresholds of $\sim$2.8 GeV and $\sim$3.6 GeV to allow prescaling of the abundant low-pT signals and thus improve the detector live-time.
 The absolute energy scale of the BEMC is calibrated in situ using high energy electrons whose momentum is determined by the TPC, while a relative tower by tower calibration is performed with more abundant minimum ionizing particles. For the preliminary $\pi^0$ cross section presented here, a subset of the available 2005 data with an integrated luminosity of $\sim 44\, \mu\mbox{b}^{-1}$ for minimum bias and $\sim 0.4\, \mbox{pb}^{-1}$ for HT triggers was analyzed. The cross section is given by
\begin{equation}
E \frac{d^3 \sigma}{dp^3} = \frac{1}{2\pi\,\Delta p_t \Delta y} \, c \, \frac{N_{\pi}}{\hat{\mathcal{L}}}
\end{equation}
where $\Delta p_t$ and $\Delta y$ are the bin widths in $p_t$ and rapidity, $\hat{\mathcal{L}}$ is the sampled luminosity, $N_\pi$ is the number of reconstructed $\pi^0$ in the bin and $c$ is an overall correction factor that contains the corrections for acceptance, reconstruction and trigger efficiency thus giving the true number of pions from the number of reconstructed pions in a given bin. These corrections are determined from a full GEANT simulation of the detector, using $p+p$ events generated with the PYTHIA event generator. Good agreement between the observed and the simulated $\gamma \gamma$ spectra has been found. 

\section{Results and Discussion}

Figure \ref{fig:XSect} shows the inclusive cross sections for neutral pions at mid- and forward rapidity and for charged pions at mid-rapidity. All cross sections are compared to NLO pQCD calculations \cite{Jager:2002xm} using the Kretzer \cite{Kretzer:2000yf} and the KKP \cite{Kniehl:2000hk} fragmentation function sets. The pQCD calculations are performed with a factorization scale $\mu = p_t$, unless stated otherwise.

Figure \ref{fig:XSect}a) shows the preliminary cross section for inclusive $\pi^0$ production in the rapidity interval $0.1 < y < 0.9$ together with the NLO pQCD calculations. Figure \ref{fig:XSect}b) shows the relative statistical errors of the data points as error bars and the preliminary point-to-point systematic errors as the shaded band. The point-to-point systematics are dominated by errors on the yield extraction from the $\gamma \gamma$ invariant mass spectra. Figure \ref{fig:XSect}c) shows the relative difference of the measured cross section to the NLO pQCD calculations for both the Kretzer and the KKP fragmentation functions. The shaded band shows the preliminary total systematic error, which is dominated by the energy scale uncertainty of the calorimeter calibration, estimated to be $\sim$5\% for the present data. The NLO calculations using the KKP fragmentation functions show excellent agreement with the data to $p_t$ below 3 GeV/c. The Kretzer fragmentation functions consistently underpredict the data. To indicate the scale uncertainty of the pQCD calculations, calculations using the KKP fragmentation functions with scales of $\mu = 2\, p_t$ and $\mu = p_t/2$ are shown by the dotted and the dashed-dotted lines, respectively.

Figure \ref{fig:XSect}d) shows the inclusive neutral pion cross section at forward rapidities in three pseudo-rapidity bins with $\left< \eta \right> = 3.3$, $\left< \eta \right> = 3.8$ and $\left< \eta \right> = 4.0$. The data points are again compared to NLO pQCD calculations using the Kretzer and the KKP fragmentation functions. At low rapidities, the KKP fragmentation functions show much better agreement with the data than the Kretzer set. At high rapidities and lower $\pi^0$ energy the data tends towards the Kretzer fragmentation functions. One consideration here is that the transverse momentum in the high rapidity bins is quite low and where pQCD at NLO might not be a good expansion. For $\eta = 4.0$ a $\pi^0$ energy of 40 GeV corresponds to $p_t\, \sim$ 1.5 GeV/c, while at $\eta = 3.3$ an energy of 40 GeV corresponds to $p_t\, \sim$ 3.0 GeV/c.

Figure \ref{fig:XSect}e) shows the inclusive charged pion cross section at mid rapidity, compared to NLO pQCD using the two different sets of fragmentation functions discussed here. Again, the KKP set shows much better agreement with the data than the Kretzer fragmentation functions. Figure \ref{fig:XSect}e) shows the ratio of the data to the NLO calculations using the KKP fragmentation functions. The excellent agreement to $p_t$ as low as 2 GeV/c is apparent. At very low $p_t$, the data tends to lie below the predictions of the KKP fragmentation and more towards the Kretzer set, as also observed for forward neutral pions. However, in the low $p_t$ regime the scale  uncertainties of the pQCD calculations are very significant, as apparent from figure \ref{fig:XSect}c), demonstrating the limited applicability of pQCD calculations in this regime of low transverse momentum.  

\vspace{-2mm}
\section{Summary and Outlook}

The presented inclusive hadron cross sections in $p+p$ collisions at $\sqrt{s}$ = 200 GeV show very good agreement with NLO pQCD calculations down to transverse momenta well below 3 GeV/c. This demonstrates the applicability of these calculations for the interpretation of measured spin asymmetries to extract the gluon polarization in the proton. All three cross section measurements presented here consistently favor the KKP fragmentation function set over the Kretzer set for $p_t >$ 2 GeV/c. However, scale uncertainties of the calculations are of comparable size to the difference between the fragmentation function sets.
Continuing analysis of the 2005 $p+p$ STAR data-set will lead to improved statistics and reduced systematic uncertainties in the inclusive $\pi^0$ cross section at mid-rapidity.


\vspace{-5mm}

\bibliographystyle{aipproc}   
\bibliography{FSimonInclusiveHadrons}

\end{document}